%% file: paper.tex
\documentclass[10pt,journal,compsoc]{IEEEtran}
\IEEEoverridecommandlockouts
\usepackage{graphicx}
\usepackage{xcolor}
\usepackage{acronym}
\usepackage{booktabs} %for top, middle and bottomline
\usepackage{multirow} %multi column and row spanning
\usepackage[space]{cite}
\usepackage{algorithm}
\usepackage[noend]{algpseudocode}
\usepackage{multirow}
\usepackage{booktabs}
\usepackage{hyperref}
\usepackage{mathtools}
\usepackage{amsmath}
\usepackage{cleveref}
\usepackage{amsmath}
\usepackage{xfrac}
\usepackage{subfigure}
\usepackage{arydshln}
\usepackage{adjustbox}
\usepackage{flushend}
\usepackage{enumitem}

\begin{document}

%\title{LOGAN: high-performance GPU-based $x$-drop alignment of long reads}
\title{The Parallelism Motifs of Genomic Data Analysis}
%\title{Exploiting GPUs to increase sequence alignment performances}
%\title{A tailored GPU design to efficiently align genome sequences}
%\title{On how to parallelize genome sequence alignment using GPUs}
%\title{Parallelizing genome sequence alignment using GPUs}
%\title{On how to Efficiently Align Genome Sequences on GPU}
%\title{On how to accelerate genome sequence alignment via GPU: the case of $X$-Drop}
%\title{$x$-drop GPU*\\

\acrodef{NW}[NW]{Needleman--Wunsch}
\acrodef{SW}[SW]{Smith--Waterman}
\acrodef{GPU}[GPU]{Graphical Processing Unit}
\acrodef{GCUPS}[GCUPS]{Giga Cell Updates Per Second}
\acrodef{GIPS}[GIPS]{Giga Instructions Per Second}
\acrodef{SM}[SM]{Streaming Multiprocessor}
\acrodef{FPGA}[FPGA]{Field Programmable Gate Array}
\acrodef{HBM}[HBM]{High Bandwidth Memory}

\author{\IEEEauthorblockN{
Katherine Yelick\IEEEauthorrefmark{1}\IEEEauthorrefmark{2},
Ayd{\i}n Bulu\c{c}\IEEEauthorrefmark{1}\IEEEauthorrefmark{2}, 
Muaaz Awan\IEEEauthorrefmark{1}, 
Ariful Azad\IEEEauthorrefmark{3}, 
Benjamin Brock\IEEEauthorrefmark{1}\IEEEauthorrefmark{2},
Rob Egan\IEEEauthorrefmark{4}, \\
Saliya Ekanayake\IEEEauthorrefmark{1}, 
Marquita Ellis\IEEEauthorrefmark{1}\IEEEauthorrefmark{2}, 
Evangelos Georganas\IEEEauthorrefmark{5}, 
Giulia Guidi\IEEEauthorrefmark{1}\IEEEauthorrefmark{2}, 
Steven Hofmeyr\IEEEauthorrefmark{1}, \\
Oguz Selvitopi\IEEEauthorrefmark{1}, 
Cristina Teodoropol\IEEEauthorrefmark{1}\IEEEauthorrefmark{2}, 
Leonid Oliker\IEEEauthorrefmark{1}}\vspace{.35cm}

\IEEEauthorblockA{
\IEEEauthorrefmark{1}Computational Research Division, Lawrence Berkeley National Laboratory, Berkeley, CA, USA \\
\IEEEauthorrefmark{2}Department of Electrical Engineering and Computer Sciences, University of California, Berkeley, CA,  USA\\
\IEEEauthorrefmark{3}School of Informatics, Computing, and Engineering, Indiana University, Bloomington, IN, USA\\
\IEEEauthorrefmark{4}DOE Joint Genome Institute, Berkeley, CA, USA\\
\IEEEauthorrefmark{5}Intel Labs, Santa Clara, CA, USA
\thanks{Extended preprint of paper ``Yelick K et al. 2020 The
parallelism motifs of genomic data analysis.
Philosophical Transactions of the Royal Society A.'' Published version to appear at
\url{http://dx.doi.org/10.1098/rsta.2019.0394} \textbf{Contact}:yelick@berkeley.edu}}}

\maketitle

\thispagestyle{plain}
\pagestyle{plain}

\input{sections/abstract}
\input{sections/introduction}
\input{sections/problems}

\input{sections/motifs}

\section*{Acknowledgments}
	This paper  summarizes work supported in part 	
	by the Department of Energy Office of Science under  
	    Lawrence Berkeley National Laboratory Contract DE-AC02-05CH11231 % Lab DEGAS, base Math, 
	and DOE contract DE-SC0008700 at UC Berkeley;  % Campus DEGAS 
        by the Exascale Computing Project
            (17-SC-20-SC), a collaborative effort of the U.S. Department
           of Energy Office of Science and the National Nuclear Security
           Administration;
         and by the National Science Foundation
	   as part of the SPX program under Award number 1823034 at UC Berkeley.
    This research used resources of the National Energy Research Scientific
         Computing Center (NERSC) under contract No. DE-AC02- 05CH11231,
    and the Oak Ridge Leadership Computing Facility under DE-AC05-00OR22725.
    both supported by the Office of Science of the U.S. Department of Energy.
        The information presented here does not necessarily 	
	reflect the position or the policy of the Government 	
	and no official	endorsement should be inferred.

\section*{Authors' Contributions}
MA, AA, BB, RE, SE, ME, EG, GG, SH, OS developed the
  parallel software described here.  KY, AB, CT and LO drafted the
  manuscript. All authors read and approved the manuscript.
  
\section*{Competing Interests}
The authors declare that they have no competing interests.

\bibliography{paper}
\bibliographystyle{IEEEtran}
\end{document}

%% file: sections/abstract.tex
% !TEX root = ../paper.tex

\begin{abstract}

Genomic data sets are growing dramatically as the cost of sequencing
continues to decline and small sequencing devices become available.
Enormous community databases store and share this data with the
research community, but some of these genomic data analysis problems
require large scale computational platforms to meet both the memory
and computational requirements.  These applications differ from
scientific simulations that dominate the workload on high end parallel
systems today and place different requirements on programming support,
software libraries, and parallel architectural design.  For example, 
they involve irregular communication patterns such as asynchronous 
updates to shared data structures. 
We consider several problems in high performance genomics analysis,
including alignment, profiling, clustering, and assembly for both
single genomes and metagenomes.  We identify some of the common
computational patterns or “motifs” that help inform parallelization
strategies and compare our motifs to some of the established
lists, arguing that at least two key patterns, sorting and hashing, are missing.   
\end{abstract}

% \begin{IEEEkeywords}
% component, formatting, style, styling, insert
% \end{IEEEkeywords}

%% file: sections/introduction.tex
% !TEX root = ../paper.tex
\section{Introduction}
\label{abs:intro}

The future of scientific computing will be increasingly data intensive due to the growth of data from sequencers, telescopes, microscopes, light sources, particle detectors, and embedded environmental sensors.  Open data policies for scientific research are leading to large community data sets of both raw and derived data.  Some of resulting data analysis problems involve massive numbers of independent computations while others require irregular computations in which the objective of the analysis is to discover the underlying structure of the data.  Many genomics problems fall into this latter category, where the structure and relationship between different sequences or entire genomes is unknown.  These problems require data structures like hash tables, histograms, graphs, and very sparse unstructured matrices.  They have dynamic sources of load imbalance and little locality, leading to unpredictable communication that is both irregular in space, with arbitrary connections between processors, and irregular in time, where one process may need data on another at any point in time.  

In this paper we describe parallelization challenges and approaches for high performance genomic data analysis using a series of examples drawn in large part from the ExaBiome project, including k-mer counting, alignment, genome assembly, protein clustering, and machine learning.  We consider analysis of both DNA and proteins expanding beyond the strict domain of genomics into proteomics. Shared memory programming is a natural fit for these problems, and indeed the most popular genome assemblers and clustering algorithms have typically run on shared memory computers. In developing High Performance Computing (HPC) implementations of these  applications, we use distributed versions of shared data structures that are updated asynchronously by individual processors with minimal global synchronization. 

\begin{figure*}
\begin{minipage}[c]{0.65\linewidth}
\begin{center}
\includegraphics[width=\textwidth]{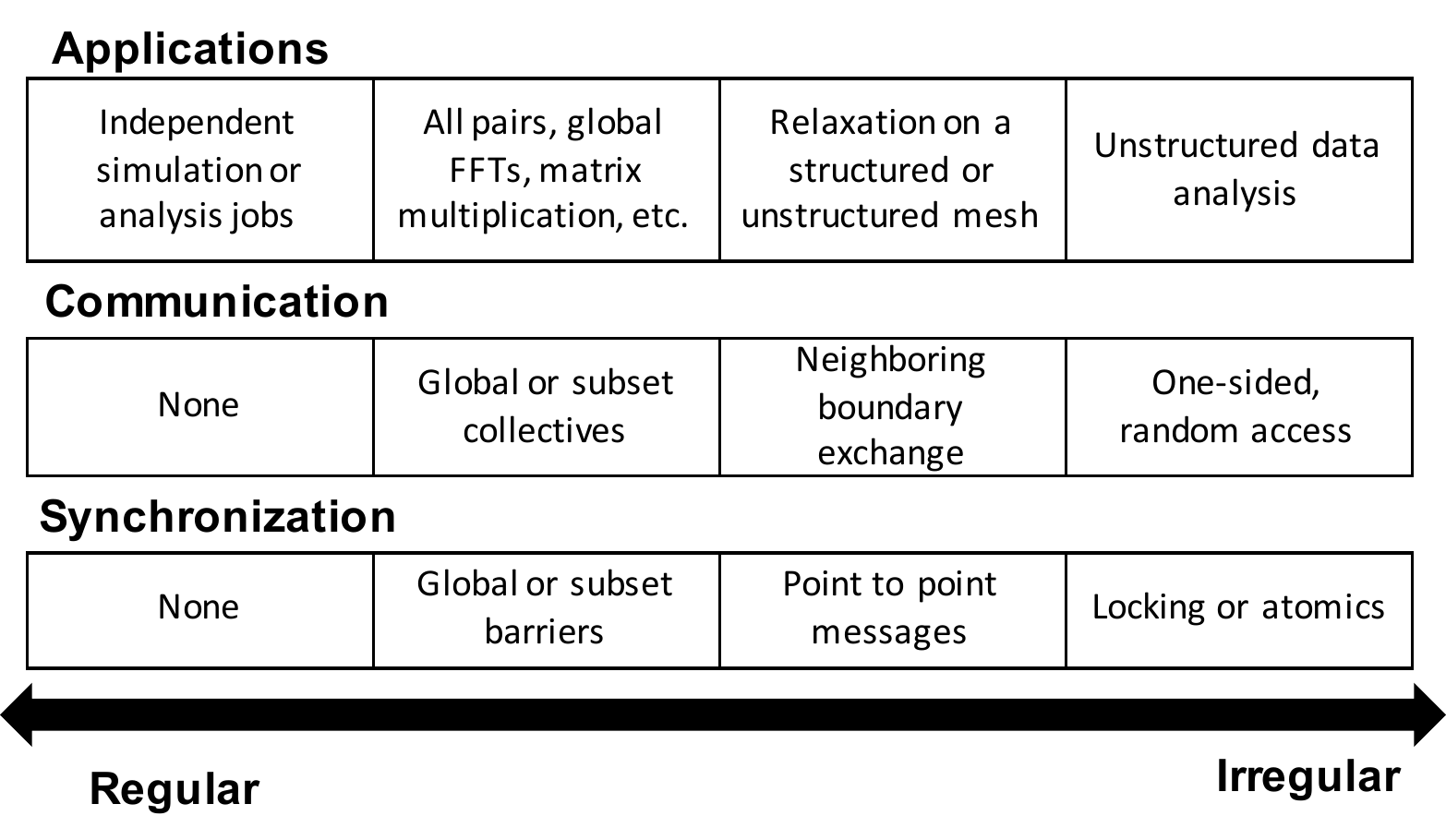}
\caption{A spectrum of regularity with different patterns of communication and synchronization.  Data analysis is often at the two extremes.}
\vspace{-0.7cm}
\label{fig:irregular}
\end{center}
\end{minipage}
\hfill
\begin{minipage}[c]{0.3275\linewidth}
     \centering
     \vspace{3.5em}\includegraphics[width=\textwidth]{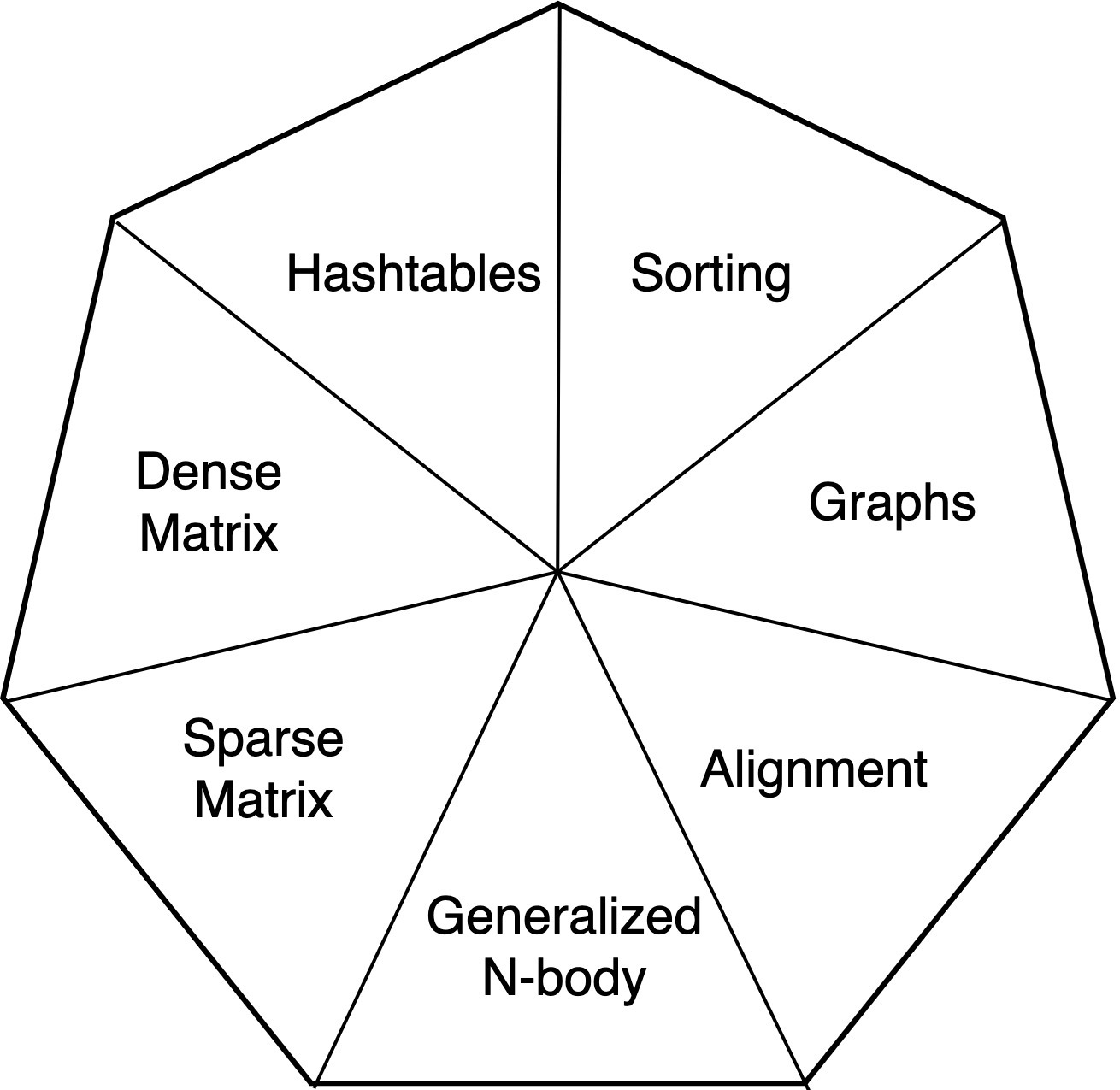}\vspace{1.5em}
     \caption{Seven parallelism motifs in genomic data analysis.}
     \label{fig:motifs}
\end{minipage}
\end{figure*}

In contrast, the applications that dominate HPC workloads are scientific simulations that have a natural degree of locality from the underlying physical laws.  These simulations often lend themselves to domain decomposition, where the physical domain is partitioned across processors, and while communication may be both global and to nearest neighbors, the presence of timesteps and iterative methods lead to natural phases of communication and computation separated by global synchronization. Figure~\ref{fig:irregular} shows a notional spectrum of simulation and analysis problems and the level of irregularity which tends to correlate with the difficulty of parallelization. On the left are independent parallel jobs, whether from simulation or analysis.  These are easily parallelized on a cluster or cloud platform using programming systems like Spark~\cite{zaharia2016apache}, or even geographically distributed computing as in the grid~\cite{shiers2007worldwide}.  Simulation problems with physical structure fall in the middle two categories, depending on whether they have global patterns of communication and synchronization, which often stress the global network bandwidth but are simpler to reason about, or involve pairwise exchange of data using synchronous or asynchronous two-sided message passing.  These boundaries are neither strict nor precise, with many applications having a mixture of styles, and deep learning landing with simulation.  However, the spectrum highlights that the genomics applications will provide an interesting perspective for the design of parallel hardware and software systems. 

In addition to summarizing parallelization techniques for genomics analysis, we identify a relatively small set of computational {\it motifs} that appear multiple times across applications, illustrated in Figure~\ref{fig:motifs}.  While we do not presume that these motifs are sufficient for all such applications, we believe they can substantively inform the design of libraries, programming systems, benchmarks, and hardware. Following a brief overview of the ExaBiome project in Section~\ref{exabiome}, Section~\ref{examples} gives an overview of several genomics data analysis problems and the computational motifs that will lead to a particular parallelization strategy.  Section~\ref{motifs} summarizes our genomics motifs and compares them to other lists of motifs, showing strong similarity to another list for data analysis problems more broadly, although we argue that two key motifs, sorting and hashing, are missing.  Section~\ref{support} describes how the motifs lead to different types of programming support and hardware requirements, and Section~\ref{conclude} makes some concluding remarks on the value and limitations of our motifs.

\section{ExaBiome Overview}\label{exabiome}

The ExaBiome project is developing scalable parallel tools to analyze microbial species, such as bacteria, fungi or viruses, which typically live in communities with hundreds of different species mixed together.  The genome level analysis of these communities, {\it metagenomics}, is key to understanding  the makeup of these communities, how they change based on external factors like temperature, moisture, or chemicals, to understand their functional behavior, and to compare them across communities. An estimated 99\% of microbial species are not culturable in isolation, making metagenome analysis the preferred technique for understanding these communities.  The human microbiome has been linked to a wide range of health issues including diabetes, cancer and mental health, while environmental microbiomes can both have both positive or negative impacts on everything from oxygen production and remediation of chemical spills to formation of toxic algal blooms. 

Exabiome, which is part of the Exascale Computing Project~\cite{messina2017exascale}, is developing HPC solutions for problems that were predominantly computed on shared memory or serial machines, and taking advantage of the processor accelerators such as Graphics Processing Units (GPUs) that are key to future exascale system designs.  The project is developing assemblers for both short and long read sequencing data (MetaHipMer and diBELLA), taking the fragmented output of sequencers and constructing long sequences from which genes, corresponding proteins, and taxonomic information can be derived. Working across large protein data sets, PISA and HipMCL extract clusters of related proteins that are useful in understanding ancestry and functional behavior. The team is also exploring deep learning techniques to relate proteins to 3D structure and function, and a set of methods to compute signatures for a metagenomes that can be used for comparisons across microbial samples in space or time and database search.

The project has already demonstrated unprecedented scales in terms of data set size and performance with the goal of growing the data set capability by more than an order of magnitude.  
The largest metagenome assembly to date used wetland soil samples that were a time series data set across several physical sites from the Twitchell Wetland in the San Francisco Bay-Delta. These samples consisted of 2.6 terabytes of 7.5 billion {\it reads}, which are the DNA fragments output from the sequencers, and the assembly computation required 5.1 hours on 1024 nodes of the Cori supercomputer at NERSC. We believe it is the largest assembly of any kind done as a single co-assembled computation, i.e., rather than pre-filtering the data in some way or assembling pieces of the data separately.  Separate analysis shows the value of such co-assemblies, especially in extracting information about the low abundance species in a sample.  The largest protein clustering computation used assembled metagenomes and metatranscriptomes from two community data sets (IMG and NCBI).  This unprecedentedly large data set contained 383 million proteins and 37 billion connections, requiring about one hour on 729 nodes of the Summit system at OLCF. 

%% file: sections/problems.tex
\section{A Sampling of Genomic Analyses}
\label{examples}

We describe at a high level some of the algorithms and parallelization approaches used in genomic data analysis, selecting a set of problems that represent a diverse set of computational patterns and are prevalent across multiple applications.  Our primary focus is on distributed memory parallelization techniques, so we describe the data distribution and communication approaches as well as any load balancing issues, or limitations to scaling when they exist.

\subsection{K-mer Analysis}

Given a set of variable-length strings, a common approach to analysing those strings is to break them into fixed-length substrings called k-mers.  For example, the string on the left has the list of 4-mers on the right.
\vspace{1em}

\begin{adjustbox}{width=\columnwidth}
\begin{tabular}{|c|}
\hline
CCTAAAGCCTA \\ 
\hline
\end{tabular}

\begin{tabular}{|c|}
\hline
CCTA\ \ CTAA\ \ TAAA\ \ AAAG\ \ AAGC\ \ AGCC\ \ GCCT\ \ CCTA  \\ \hline
\end{tabular}
\end{adjustbox}

\vspace{1em}
Several bioinformatics analyses involve counting the number of occurrences of each distinct k-mer, e.g., to filter low-frequency k-mers that are likely errors, to find high frequency k-mers that indicate repetitive regions of the genome, or by using the k-mer histogram as a signature for a set of genomics data. K-mers also serve as seeds in determining whether two DNA fragments are likely to align with one another and may also be used on protein data with its 21-character amino acid alphabet in addition to DNA.   The most common approach to k-mer counting is to build a hash table of k-mers, possibly using a Bloom filter, an approximate and space efficient data structure that answers queries about set membership,
to eliminate singletons.  If the k-mer length is small, a direct map may be practical, and sorting is also possible, although to keep memory use in check, k-mers are generated incrementally and identical ones merged while they are sorted to avoid having all of them in memory at once. Manekar and Sathe give an overview of the various approaches and benchmark some of the popular shared memory tools~\cite{manekar2018benchmark}.

Distributed memory parallelism for k-mer counting becomes increasingly important to address large sets of environmental microbial genomes and to handle cross-genome comparisons.  Raw sequencing data may be several times larger than the final genome, e.g., the data set may be sequenced repeatedly giving it a sequence {\it depth} of 10-50x to ensure that every location is sequenced multiple times so that sequencing errors can be eliminated.  Large environmental data sets may therefore run to multiple terabytes resulting in input data that does not fit on a single large-memory compute node. The list of k-mers --- prior to removing duplicates --- requires storage nearly {\it k} times larger than the input, and given that typical k-mer lengths may  run from 10-50 characters, the raw k-mers may not fit even in the aggregate memory of a multi-node system.  The ExaBiome project uses a hash-based approach to k-mer counting with a Bloom filter used to avoid storing most of the singleton k-mers.  The resulting hash table stores the count of each unique k-mer that occurs more than once in the original data. 

In general, there is no communication locality when building the distributed hash table, so on p processing nodes, each k-mer will be communicated remotely with probability (p-1)/p.  This creates an irregular many-to-many communication pattern without any predetermined patterns and without natural points for bulk-synchronous communication. A Bloom filter is useful to avoid storing singleton k-mers, but requires the same irregular many-to-many communication as the hash table: all k-mers are communicated, but the Bloom filter requires only a few bits for each unique k-mer. A good hash function can ensure even load balance of the unique k-mers, but significant communication load imbalance still results when the frequency distribution is skewed, as is often the case in real data sets where there are some very high frequency k-mers. Local aggregation of such ``heavy hitters'' can reduce communication bottlenecks for high frequency k-mers, but the effectiveness depends on having a small number of such k-mers so that a local table can collect and combine them.

Within the ExaBiome project we have multiple instances of k-mer analysis, which include a basic count / histogram operation and indexing to collect the information about the position of each k-mer in the set of input sequences (reads). Memory utilization is a key factor in design, and to avoid having the full list of k-mers (with duplicates) in memory at any given point in time, one version of the code performs all-to-all exchanges in phases~\cite{georganas2014parallel,georganas2015hipmer, georganas2016scalable, georganas2018extreme} and counts k-mers for use in the short read assembly, and another version keeps indexing information for use in computing long read overlaps~\cite{ellis2019dibella}. Other distributed memory k-mer analysis tools include Bloomfish, which uses a similar MPI all-to-all collective approach but has only a single phase~\cite{gao2017bloomfish} thus limiting data set size due to memory constraints, and Kmerind, which has demonstrated scaling to over 20 TB data sets by using multiple phases and various memory saving optimizations~\cite{pan2017kmerind}.  The most recent ExaBiome k-mer counting tool is entirely without global synchronization and uses one-sided communication to continually send k-mers while combining and storing local ones. Not only does it avoid global synchronization, but it also hides latency by using non-blocking communication. 

\subsection {Pairwise Alignment}
Alignment is performed on both DNA and proteins to find approximate matches between strings, allowing for a limited number of insertions, deletions, and substitutions.   
Pairwise alignment is typically done with some form of dynamic programming, i.e., Needleman-Wunsch~\cite{needleman1970general} 
for the best overall alignment or Smith-Waterman~\cite{smith1981identification} for the best local substring alignment.  
Both algorithms find an optimal match based on a given scoring scheme that rewards matches and penalizes mismatches, insertions, and deletions. The algorithms operate by filling in an $n \times m$ scoring matrix based on strings of length $n$ and $m$ and compute the optimal score at each position with an overall sequential cost of $O(n m)$. The resulting dependence pattern leads to parallelism along an anti-diagonal wave-front. 
A popular heuristic algorithm, called X-drop~\cite{zhang2000greedy}, searches only for high-quality alignments by tracking the running highest score and not exploring cell neighborhoods in the matrix whose score drops by a given threshold below the maximum. 
It gets its performance benefits from dynamically resizing the anti-diagonal wave-front (i.e., its {\it band}), therefore reducing the search space, and may stop early when there is no high-quality match.  

Pairwise alignment appears throughout genomic data analysis, because both errors in data from sequencers and variations in genomes across individuals lead to imperfect string matches.  The ExaBiome project has multiple instances of alignment, which include aligning short reads to partially assembled sequence data (called {\it contigs}), aligning long reads to each other, or aligning proteins to each other.  
Typical lengths of DNA from sequencers run from 100-250 characters for short reads to over 10,000 for long-read technology reads, while proteins are typically a few thousand characters long.  Even if one is aligning against a full genome, e.g., the 3-billion-character reference human genome or a large database of genomes, it will be done by starting from a predetermined location or seed as described in the next section.  At the scale of a few hundred to a few thousand characters, pairwise alignment is amenable to SIMD~\cite{farrar2006striped, li2018minimap2, suzuki2018introducing}, multicore, GPU~\cite{liu2013cudasw++, mr-cudasw2014, FengIcpp2019, logan2020}, and even FPGA~\cite{li2007160,di2017architectural} parallelism, and can take advantage of narrow data types to represent the four nucleotides in DNA, the twenty-one amino acids in proteins, or the limited range of values in the scoring matrix. Recent work also shows how dynamic programming problems exhibit essentially linear speedups using the concept of rank convergence~\cite{maleki2016efficient}, in which the pairwise alignment is computed via a series of dense matrix multiplications on the tropical semiring where the scalar addition is replaced with the maximum operator and scalar multiplication becomes integer addition.

Alignment dominates the local on-node computation in ExaBiome applications, as well as other genome analysis tools across scales.  However, there is not sufficient work for distributed memory parallelism within pairwise alignment, and even GPU offload requires batch alignment, where a set of pairs are aligned as a single operation, to amortize the startup and data movement overhead.

\subsection{All-to-All Alignment}
\label{sec:many2many}
Alignment is often done across a set of strings, such as alignment against a database of reference genomes or proteins, a set of patient genomes against a single (large) reference, or a set of reads from a sequencer against each other or against partially constructed genomes fragments as part of genome assembly.  The ExaBiome project performs all-to-all alignments as part of short read assembly (merAligner within MetaHipMer), in which case the input reads are aligned against all partially assembled contigs~\cite{georganas2015meraligner}, and as the first step in long-read assembly where reads are aligned against each other in BELLA and diBELLA~\cite{guidi2018bella,ellis2019dibella}.

The all-to-all computational pattern is familiar from n-body simulations and, as in that case, computation on all O($n^2$) pairs of strings/particles is prohibitively expensive.  To tackle this, particle simulations rely on hierarchical tree-based approaches that exploit the physical layout of particles in space, which is not applicable in alignment.  Instead, in aligning a set of sequences, one can pre-filter the pairs to find ones that are likely to have a good alignment. Our approach therefore looks for sequences that share at least one short identical string, e.g., a k-mer, which can also be used to seed the alignment.  For example, to align a set $S$ against another $T$, store all k-mers from strings in set $T$ in a hash table and lookup all the k-mers from strings in $S$ to find matching pairs, starting each pairwise alignment from the position of the common k-mer.   

In distributed memory, the k-mer hash table has an irregular many-to-many communication pattern that is familiar from the k-mer counting, but each k-mer now retains the list of sequences containing that k-mer.  
The hash table may be viewed as a sparse k-mer$\times$sequence matrix with sequences from set T.  
To compute the set of sequences from S that have a matching k-mer, we can take either a linear algebra or database “hash-join” view of the problem.  

In the former case, we construct a k-mer$\times$sequence matrix for each set, transpose one and multiply them to obtain a sparse sequence$\times$sequence where each nonzero at position $i,j$  represents a pair $S_i,T_j$ that share a common k-mer. The sparse matrix primitive that performs this operation is known as SpGEMM, for Sparse GEneralized Matrix-Matrix multiplication~\cite{bulucc2012parallel}. It is generalized in the sense that the multiplication can operate on any arbitrary algebraic structure, also known as a semiring, and not just the real field.
The single-node shared memory BELLA code uses this approach~\cite{guidi2018bella} to align a set of long reads to itself, so $S = T$. Both input and output matrices in BELLA's case are sparse.

The second approach constructs the same k-mer$\times$sequence table for $T$ but does not explicitly compute the sequence$\times$sequence matrix.
Instead, as it computes the set of k-mers in $S$, it looks them up in $T's$ table to find sequences in $T$ with a common k-mer.
The distributed memory diBELLA uses this approach in a bulk-synchronous series of many-to-many exchanges, while merAligner performs alignments on-the-fly as the read sequences are processed (typically fetching the contig from a remote processor) that contains a matching k-mer. merAligner also caches these contigs as there is enough likely reuse that can be leveraged to save repeated communication of contigs.  

All of these distributed memory alignment algorithms involved irregular many-to-many communication either done asynchronously as 1-sided remote look-ups or in batches.  
The asynchronous approach has more messages, each of which is small, so communication software overhead and latency can limit performance.  It has the advantage of overlapping computation and communication together, which makes good use of both networking and computing resources.  The bulk-synchronous approach leads to better message aggregation between pairs of processors, but it can suffer from high load imbalance costs due to the implied barriers at each exchange. However, separating communication from computation prevents overlap and is more likely to trigger bisection bandwidth limits in the network.  
The pairwise alignments that follow communication can either be done one pair at at time or in batches, with the bulk-synchronous version likely having larger batches to do.

K-mer-based matching is not the only method that is used to index large genomic data sets. In particular, suffix trees and their more practical sibling suffix arrays provide an alternative way of indexing large data sets.  Rather than hashing, these methods using sorting and search on a compact representation of the suffix substrings and then build a hierarchical index representation of the data. Suffix arrays are significantly more flexible than direct k-mer based approaches because they effectively index all possible k-mer lengths at once. However, they are harder to implement and they often come with increased computational costs. Recent work on distributed suffix array construction~\cite{flick2015parallel} as well as querying~\cite{flick2019distributed} has shown scaling to eight nodes but with the potential to make these data structures more popular in HPC approaches.

 There are other applications that arise in comparing which genomes or metagenomes align to each other. 
In this scenario, one is often interested in some sort of ``distance'' metric between pairs of genomes or metagenomes, as opposed to merely identifying the candidate pairs that might align. The output is often dense because almost all pairs of (meta)genomes will contain conserved regions that will provide a match using shared k-mers. Using an approach similar to BELLA, Besta et al.~\cite{besta2019communication} use parallel sparse matrix computations to compute the Jaccard similarity between all pairs of genomes. They also utilize the aforementioned SpGEMM primitive, with one difference that the software is optimized for the case where the output genomes$\times$genomes matrix is dense because it holds the Jaccard similarity.
 
The Bioinformatics community have been developing alternative space-efficient data structures in order to compute (meta)genome-to-(meta)genome distances for the scenario where a distributed-memory computer is unavailable. MASH~\cite{ondov2016mash}, perhaps the most popular of such tools, uses the MinHash sketch technique~\cite{indyk1998approximate} for each (meta)genome and only computes the Jaccard similarity on those sketches, as opposed to finding explicit shared k-mers. Recently, Baker and Langmead~\cite{baker2019dashing} took the sketching approach one step further and used the HyperLogLog (HLL) algorithm for further compression. While we are not aware of any distributed-memory approaches to sketch-based genomic distance calculations, the HLL data structure itself is trivially mergeable. HLL has been utilized in distributed genome assembly for efficient k-mer counting in the past~\cite{georganas2014parallel}. We therefore expect forthcoming developments in distributed-memory sketch based genome comparison.

\subsection{Graph Traversal for Genome Assembly}
\label{graph}

Genome assembly involves the  analysis of reads from sequencers to produce longer contiguous sequences of the  genome with errors  corrected.  For short reads with their low error rate ($<.1\%$),  the  MetaHipMer software performs k-mer analysis and eliminates low frequency k-mers which are presumably errors.  Along with each k-mer in the final hash table, it  stores left and right {\it high  quality} extensions, i.e., the character that frequently appeared to the left and right of the k-mer in the original input.  This table is then viewed as a De Bruijn~\cite{de1946combinatorial} graph in which a k-mer vertex is connected to another if their k-mers overlap in k-1 contiguous positions.  The left and right extensions with each k-mer make it straightforward to find neighboring vertices.  A depth-first traversal starting from arbitrary k-mers compute the connected components of the graph which are linear sequences called {\it contigs}.   For metagenomes, the same basic method is used but with increasing values of $k$, with contigs from the earlier steps added as reads to the later ones.  This iterative process helps to improve coverage of low-depth, highly fragmented genomes in the earlier phases and resolve repeated regions and obtain longer contigs in the later phases.   Once the contigs are formed, the assembler builds a graph with contig vertices and uses alignment to find reads that align to multiple contigs and thus form an edge in the contig graph.  There are several other graphs traversals performed on both the k-mer and contig graph, which are omitted here.  We focus on parallelization of contig construction on the k-mer hash table.  A more detailed description of contig generation and other graph traversals during assembly are available in the HipMer and MetaHipMer papers~\cite{georganas2014parallel,georganas2015hipmer,georganas2016scalable,georganas201718,georganas2018extreme}.

MetaHipMer takes advantage of the memory and computing performance of distributed memory supercomputers to support large-scale assemblies.   The hash tables involved in our algorithms can be up to tens of terabytes and do not fit in a typical shared memory node, and contig generation is written in UPC~\cite{carlson1999introduction,upc2005unified} so that hash table buckets are directly accessed by any processor using one-sided communication.  During construction, we aggregate multiple insert operations intended for the same remote processor to amortize communication overhead.  This is done dynamically and asynchronously: once a particular buffer for a remote node is full, it is sent using one-sided memory operations with atomics to the memory of a remote processor.  Hash table inserts and lookups are done in two separate phases, so the delayed inserts from aggregation are not semantically visible --- all of the inserts are complete at the end of the phase and the order is not important. 

During graph traversal the hash table remains fixed, although multiple traversals happen in parallel from different starting vertices and individual k-mer vertices are marked as visited to avoid duplicate traversals.  This is done with fine-grained remote atomics rather than locking to minimize the number of communication round trips, although this stage is latency-limited since each processor is performing a single-threaded traversal of the graph and needs to wait for a remote vertex before continuing.  In later stages of assembly, the hash table of contigs is truly read-only and each contig may be used multiple times by a single processor, so caching remote contigs is efficient and preserves correctness.  
Caching is not performed during contig generation because there is limited reuse.

\subsection{Sparse Matrix Operations for Protein Clustering}

Proteins of the same evolutionary origin are said to be homologous. Homologous proteins often perform similar functions; hence homology finding facilitates protein annotation and the discovery of novel protein families. One often infers homology from excess sequence similarity; with ``excess'' referring to higher similarity than can be encountered by chance. Even then, a simple pairwise similarity metric is just a proxy for homology and can lead to both false positives and false negatives, depending on the parameters used in sequence similarity calculations.  A clustering step that takes the similarity matrix as input and exploits topology information (i.e., the transitivity of neighboring proteins) to find more robust and accurate protein families. This helps eliminate a significant portion of spurious homology connections and recovers many missing links while computing a globally consistent view of the clusters. 

A typical pipeline for protein clustering therefore involves first finding highly-similar sequences using many-to-many alignments among proteins, using one of the popular tools such as MMseqs2~\cite{steinegger2017mmseqs2}, DIAMOND~\cite{buchfink2015fast}, or LAST~\cite{kielbasa2011adaptive}. K-mer based indexing that is similar in spirit to those described in Section~\ref{sec:many2many} is often used to reduce the number of comparisons. The ExaBiome project is currently working on a novel many-to-many protein similarity search tool, tentative called Protein Sequence Aligner (PISA), that is scalable to Exascale architectures. The result of the similarity matrix/graph computation is then fed into a clustering algorithm that discovers the ultimate protein families. Since this two-step process is often very expensive, single-step clustering algorithms~\cite{li2006cd} have gained in popularity among those who does not have access to high-end computing equipment, despite often resulting in fragmented clusters. We will not be focusing on those methods here because one of the goals of the ExaBiome project is to improve accuracy by utilizing Exascale computers. 

The Markov Cluster (MCL) algorithm~\cite{enright2002efficient} is arguably the canonical graph-based algorithm for clustering protein similarity matrices. The MCL algorithm treats this similarity matrix as an adjacency matrix of the graph where vertices are proteins and edges are similarities. The graph is sparse because only those similarities that are above a certain similarity threshold are retained. MCL performs random walks from every vertex (protein) in the graph. It exploits the fact that most of these walks will be trapped within tightly connected clusters, hence driving up the probability mass that is accumulated within each cluster. In order to avoid densifying the intermediate matrices and making the computation infeasible, MCL performs various pruning strategies that are shown to not hurt the quality of the final clusters~\cite{van2008graph}. 

The simultaneous random walks directly map to a sparse matrix primitive that is commonly known as SpGEMM, which computes the product of two sparse matrices. The high-performance distributed re-implementation of the Markov Cluster algorithm, known as HipMCL~\cite{azad2018hipmcl}, utilizes some of the most general and scalable sparse matrix algorithms implemented within the Combinatorial BLAS~\cite{bulucc2011combinatorial}. These algorithms include a 2D SpGEMM algorithm known as Sparse SUMMA~\cite{bulucc2012parallel}, several different shared memory SpGEMM algorithms~\cite{nagasaka2019performance} that are optimized for different iterations of HipMCL, a fast memory estimator based on sparse matrix dense matrix multiplication for memory-efficient SpGEMM~\cite{Cohen1998}, as well as a very fast distributed memory connected components algorithm~\cite{azad2019lacc} that is used for extracting the final clusters from the result of the HipMCL iterations. The integration of GPU support to HipMCL as well as other performance improvements for pre-exascale architectures has recently been published~\cite{hipmcl2}. Using faster communication-avoiding SpGEMM algorithms~\cite{azad2016exploiting} for HipMCL is ongoing work. 

\subsection{Machine Learning for Genomics and Proteomics}
\label{sec:ml}
\begin{figure}
\begin{center}
\includegraphics[width=\columnwidth]{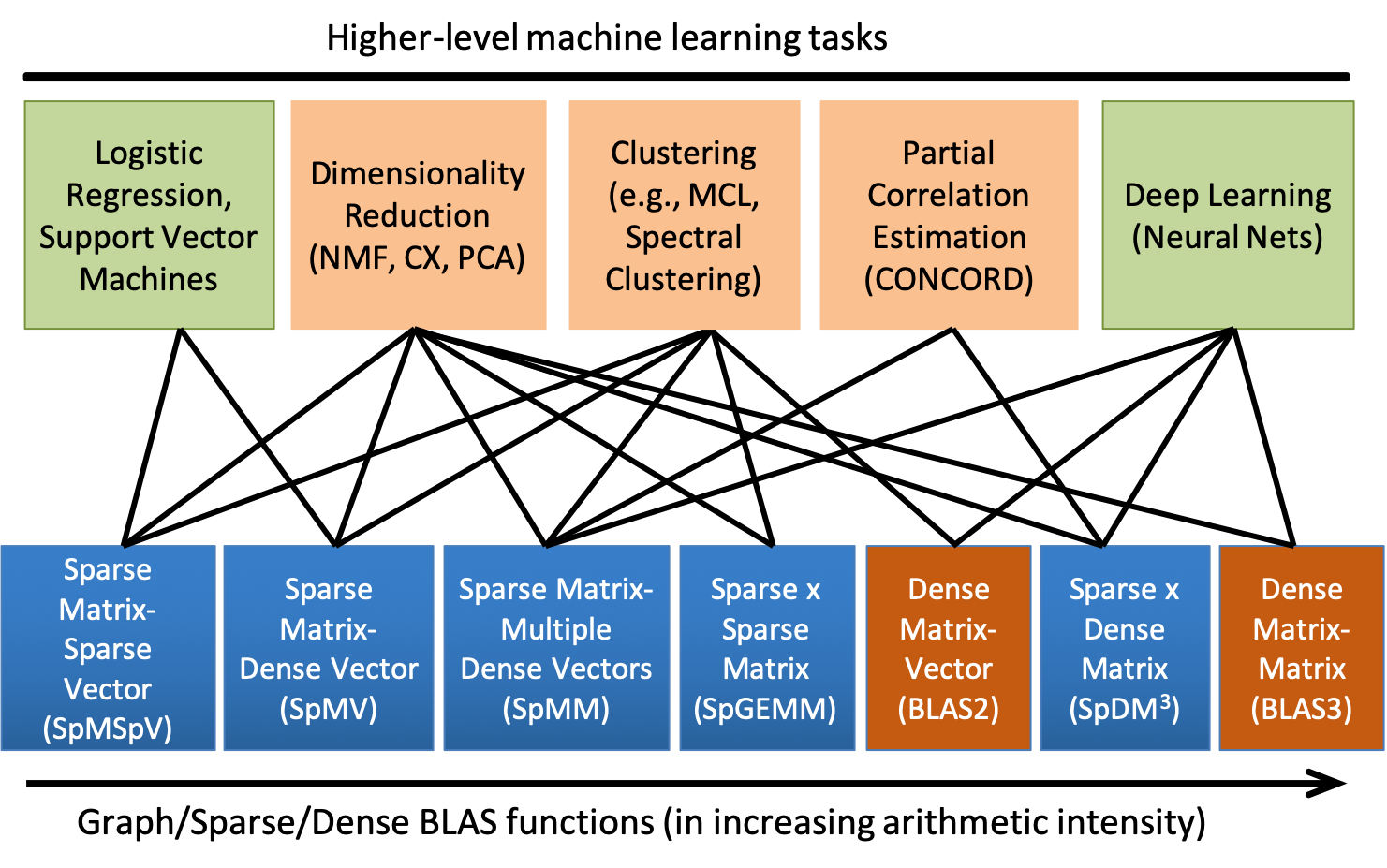}
\caption{Dependencies of various machine learning methods upon linear algebraic primitives. Orange boxes are unsupervised methods whereas the green boxes include supervised methods. NMF stands for non-negative matrix factorization, PCA stands for principal component analysis, and MCL stands for the Markov cluster algorithm. CONCORD stands for CONvex CORrelation selection methoD. CX refers to a low-rank matrix factorization in which a subset of the columns (C) is one of the factors. }

\label{fig:mldiagram}
\end{center}
\end{figure}

A comprehensive coverage of machine learning (ML) applications in genomics and proteomics is both too large and too fast growing to address here. Instead, we touch on the computational building blocks for the machine learning algorithms that are commonly applied to genomic and proteomic data.
A large class of machine learning methods are built on top of basic linear algebraic subroutines that are found in the modern dense BLAS~\cite{dongarra1990algorithm}, Sparse BLAS~\cite{duff2002overview}, or the GraphBLAS~\cite{graphblasapiC}. This relationship is illustrated in Figure~\ref{fig:mldiagram}.

Machine learning has been applied to metagenome assembly in various contexts. For example, MetaVelvet-SL~\cite{sato2014metavelvet} uses Support Vector Machines (SVMs) to identify the potentially chimeric nodes on a metagenomic de Bruijn graph. A chimeric node is shared by the genomes of two closely related species and needs to be split into multiple nodes for an accurate assembly. 
A popular application of ML to proteomics data is to discover ancestral relationships (i.e. homology) between proteins. Kernel-based methods, such as SVMs, have been traditionally applied to this problem~\cite{rangwala2005profile}.
Other fundamental problems in this domain are protein folding~\cite{dill2008protein}, especially the prediction of the 3D structure of the protein~\cite{alquraishi2019end}, and protein function prediction. The function of a protein can be predicted using either the sequence, the 3D structure of the protein, or both~\cite{gligorijevic2019structure}.

% This paragraph is not in the journal version
Recently, there has been a growing set of deep learning (DL) approaches to these problems. The particular DL machinery is input and problem specific but include transformers used in language modeling~\cite{devlin2018bert}, convolutional neural networks (CNNs), and graph neural networks (GNNs). These different DL machines have different computational footprints: Transformers heavily rely on relatively large dense matrix computations, CNNs can be trained using a series of small matrix multiplications~\cite{georganas2018anatomy}, and GNNs are bottlenecked with large sparse matrix-dense matrix multiplications~\cite{kipf2016semi}.

%% file: sections/motifs.tex
\section{Comparison with other Parallelism Motifs}
\label{motifs}
Several computational patterns arise in the ExaBiome application and are common to other genomics
applications and data analysis problems more broadly.  These are displayed in Figure~\ref{fig:motifs} and include:

\begin{enumerate}[labelindent=\parindent,leftmargin=\parindent]
  \item Hash tables: These are used throughout the genome assembly applications, MetaHipMer and diBELLA, to store
    k-mers for the purposes of counting (histogramming), and for quickly finding pairs of sequences
    with a common substring.
  \item Sorting: While used less frequently than hashing in our examples, it is another technique
    for counting k-mers and is used in suffix arrays and to prioritize graph operations, e.g., finding the longest contig
    as a starting point for a graph traversal.
  \item Graph traversals: Used to connect k-mers into contigs and in other analyses on the contig
    graph to resolve ambiguities and increase assembly length.
  \item Alignment: The problem of finding the minimum edits required to make two strings match
    is used on raw read data, assembled genomes, genes, and proteins.
  \item Generalized n-body: The problem of comparing or aligning all sequences in one set to another
    set (or the same set), but using a method such as limiting to pairs with a common k-mer to
    avoid all $O(n^2)$ comparisons. 
  \item Sparse matrices: Sparse matrix products used within the generalized n-body problem to find
    pairs, for protein clustering, etc.
  \item Dense matrices: There is ongoing work by ourselves and others to use
    machine learning methods in genomic data, which often make use of dense matrix multiplication 
    as described in Section~\ref{sec:ml}. Further, pairwise alignment can in theory also be computed 
    using dense matrix computations on semirings. 
  \end{enumerate}

In addition to these seven motifs of genomic data, local computations such as parsing
reads into k-mers and other basic string operations, arithmetic operations, logical operations,
and more occur in all of our applications.  When these can be performed independently on separate
data, they can be invaluable in obtaining high performance parallel implementations, but if the
operations are each performed serially they are not instrumental in understanding parallelization.
In comparing to other lists of motifs, we include these as "Basic Operations" in Table~\ref{table:motifs} although they tend to 
be linear time operations on the input which can be almost trivial to parallelize, and thus less useful as a 
parallelism motif.

While our selection of problems informing our genomics motifs is naturally biased, we note that independent HPC researchers have been focusing on similar problems. For example, Darwin~\cite{turakhia2018darwin} is a co-processor specifically designed to perform fast all-to-all long read overlapping and alignment in the context of assembly. The body of work from Aluru's 
group at Georgia Tech similarly encompasses k-mer analysis, alignment, assembly, and clustering and uses some of the same patterns albeit pushing in the direction of bulk-synchronous computation~\cite{aluru2016genomes}. SARVAVID~\cite{mahadik2016sarvavid} provides a Domain-Specific Language (DSL) with language constructs for k-mer extraction, index generation and look-up, clustering, all-to-all similarity computation, graph construction and traversal for genome assembly, and filtering error-prone reads. Mahadik et al. identify this list of "kernels" as common to a broad variety of genomics applications. We remark that these kernels can be mapped to our own list of motifs.

There are other proposals for the parallelism motifs that cover many applications of scientific
simulations, data analysis, and more.  The original set of ``Seven Dwarfs'' due to Phil
Colella~\cite{colella2004defining} was meant to capture the most important computational patterns in
scientific simulations and are shown in the first column in Table~\ref{table:motifs}.  The Berkeley
View report~\cite{asanovic2006landscape} on multicore parallelism, in the second column, generalized
these patterns to capture a broader set of applications including some data analysis problems.  A report by
the National Academies~\cite{national2013frontiers} then defined a set of ``Seven Giants'' of Big
Data, shown in the third column, which combined sparse and dense matrices into a single motif. 
Ogres~\cite{fox2015towards} is another similar, yet multidimensional classification of both HPC and Big Data applications
based on 51 well-studied NIST applications.
Our own genomics motifs in the
last column are quite similar to those in the ``Seven Giants'' set, but in our view the ideas of hashing and sorting are so essential to
understanding data analysis for genomic data and for other large-scale database analyses involving 
joins that they deserve to be separate categories. 
They are also standard in other large-scale database operations.
On the other hand, optimization and integration are very general techniques that can lead to a
variety of parallelism patterns depending on the data and method being used, e.g., they may be
dominated by dense or sparse matrix operations, as well as other independent computations.  Each
list takes a somewhat different approach to characterizing independent operations, which in our view
is such a general notion that it does not belong as an algorithmic motif.  Colella's Monte Carlo
class is a more specific class of problems that do lead to a style of parallelism, albeit dominated
by independent calculations.

\begin{table}
\caption{A comparison of motifs for parallel computing, including our own set for genomic data analysis. *Basic operations include string parsing, string identity, and 2-bit encoding of DNA sequences.}
\adjustbox{width=\columnwidth}{
\begin{tabular}{l|l|l|l}
    \toprule
    {\bf Colella 7 Dwarfs} & {\bf Berkeley View Motifs} & {\bf NRC 7 Giants} & {\bf Genomics Motifs} \\
    \hline
    Dense Matrix	& Dense Matrix	& Dense and 	    & Dense Matrix \\ \hline
    Sparse Matrix	& Sparse Matrix	& … Sparse Matrix	& Sparse Matrix \\ 
    \hline
    Structured Grid & Structured Grid	&   &	\\ 
    \hline
    Unstructured Grid	& Unstructured Grid &    &  \\ 
    \hline
    Spectral Methods	& Spectral Methods	 &  &   \\ 
    \hline
    Particle Methods	& N-Body	& Gen. N-Body	& Gen. N-Body \\   
    \hline
    Monte Carlo 	& MapReduce	& Basic Statistics	& Basic Operations*  \\  
    \hline
    & Finite State Machine	& & \\ 
    \hline
    & Graph Traversal	& Graph Theoretic	& Graph Traversal  \\ 
    \hline
    & Dynamic Prog.	& Alignment	& Alignment \\
    \hline
    & Backtracking Search	& & \\ 
    \hline
    & Graphical Models	& & \\
    \hline
    & Combinatorial		& & \\ 
    \hline
    & & Optimization	   & \\ 
    \hline
    & &	Integration 	& \\ 
    \hline
    & & &	Hash Tables \\
    \hline
    & & &	Sorting  \\ 
    \bottomrule
\end{tabular}
}
\label{table:motifs}
\end{table}

\section{Hardware and Software Support for Parallel Genome Analysis}
\label{support}

Although some analysis problems can be done independently or with
traditional bulk-synchronous parallelism, we argue that the irregular
and asynchronous nature of some of these problems~\cite{georganas2016scalable, georganas2017merbench, ellis2017performance} places different
requirements on the programming systems, libraries and network than most simulation problems. In addition, communication optimizations 
have a somewhat different characteristic than in more structured and regular computations.

Roughly speaking, there are four programming styles for distributed memory communication:
\begin{itemize}[labelindent=\parindent,leftmargin=\parindent]
  \item Bulk-synchronous collectives, such as broadcast, reductions,
    and all-to-all exchanges.  For example, MPI collectives have a rich set of
    collective operations~\cite{gropp1999using}.
  \item Two-sided point-to-point communication, i.e., send and  receive, which 
   need not be synchronous, but requires two-sided coordination and is,
   therefore, often done in bulk-synchronous phases.  MPI is again the
   standard here with various forms of send and receive.
   \item One-sided shared memory or Remote Data Memory Access  (RDMA),
     including put, get, and atomic memory operations.  There are
     several examples languages that support this style, including UPC
     used in the original MetaHipMer assembler~\cite{carlson1999introduction}.
   \item Remote Procedure Call (RPC), which invoke remote computation
     while communicating input and output arguments between
     processors.  The most recent version of UPC++ provides a set of RPC features with
     asynchrony-by-default to encourage communication overlap~\cite{bachan2017upc++}.  
  \end{itemize}

The majority of simulation codes are written in some combination of
the first two styles, while data analytics problems written in a
map-reduce framework use collectives.  But for analytics problems
involving hash tables with random-in-time and random-in-location
access, we argue that the latter two are a  better fit.  Sparse matrix computations such as iterative
methods can be programmed elegantly using bulk-synchronous
parallelism, as can sorting and generalized all-to-all problems, although
the data exchanges are often irregular and unbalanced, with
the volume of data between processors varying considerably. 
Communication imbalance issues can affect sparse matrix multiplication when the 
distribution of nonzeros is nonuniform, e.g., when a k-mer appears in 
many of the input sequences, or in parallel sorting when the 
distribution of values being sorted is nonuniform, e.g., a single 
value appears with very high frequency.  These imbalance factors may encourage 
designs that avoid global communication and synchronization in favor of 
overlapped point-to-point or one-sided communication. 

While numerical libraries form  the basis of many computational
simulations, we see distributed data abstractions for hash  tables,
Bloom filters, histograms, and various types of queues for rebalancing
data and computational load as keys to our analysis problems.  For
example, the  Berkeley Container Library~\cite{brock2019bcl} provides the data
structures and CombBLAS provides the distributed memory sparse matrix
primitives designed for graph algorithms~\cite{bulucc2011combinatorial}.
These libraries can capture some of the more important communication optimizations, 
which are familiar ideas but have somewhat different usage. 
\begin{itemize}[labelindent=\parindent,leftmargin=\parindent]
    \item Asynchronous communication avoids both global and pairwise synchronization, 
    allowing each thread to progress without waiting to resolve load imbalance from 
    communication or computation that may vary over time. 
    \item Non-blocking communication provides overlap for both 
    computation and other communication events, and is especially important 
    for fine-grained communication to avoid paying full latency costs for each 
    message. In a one-sided model this means non-blocking put and get operations
    or fire-and-forget in an RPC model.
    \item Communication aggregation is a standard technique in bulk-synchronous applications, 
    but in asynchronous ones this involved dynamic buffering of data destined for a single 
    core or node and shipping it when the individual buffer is full or based on some other trigger.
    In practice the management of the message buffers creates a critical trade-off between 
    memory footprint and number of messages, but the uncertainty of communication volume and 
    destination makes this particularly challenging. 
    \item Improving spatial locality is not always possible for irregular data, e.g., 
    hash table construction on unknown data, but when insight into the data is possible, a carefully constructed hash function can provides significant benefit in reducing
    the percentage of remote accesses~\cite{georganas2015meraligner}. 
    \item Caching remote data is useful when there is sufficient temporal locality, e.g., 
    in looking up contigs during alignment of reads to contigs during assembly.
    \item Iteration space tiling used in communication-avoiding algorithms for dense 
    matrix multiplication\cite{agarwal1995three,solomonik2011communication} and 
    n-body calculations~\cite{hendrickson1995parallel,driscoll2013communication} provide provable advantages in 
    reducing communication volume and number of messages at the cost of additional memory. These methods do not 
    simply partition the result matrix or particles/sequences over processors, but instead replicate them
    to the extent allowed by available memory. For sparse matrices
    and sparse interactions the benefits depend more on the sparsity 
    patterns~\cite{Cohen1998,bulucc2012parallel,ballard2013communication,nagasaka2019performance}, 
    but are useful in clustering~\cite{azad2018hipmcl} and possibly alignment.
\end{itemize}

From an architectural perspective, these highly irregular applications
stress message injection rate, communication latency, and in some
cases bisection bandwidth~\cite{georganas2016scalable, georganas2017merbench}. 
They may never saturate link bandwidth if
a multi-core node cannot inject small messages into the network fast
enough to saturate bandwidth.   While message aggregation is used in
our implementations to maximize bandwidth utilization, this tends to
put significant pressure on the memory per node due to the nearly
random pattern of remote processors with which a single node communicates.
Communication overlap can also be critical in these applications,
including overlapping multiple communication events with each other.
Perhaps the most obvious difference between genome analysis and
simulation is that floating point numbers are essentially nonexistent
in the lower level analyses and only arise in machine learning such as
clustering and deep learning application.

\section{Summary}
\label{conclude}

This papers provides an overview of some of the computational patterns that arise in
genomics data analysis using examples from the ExaBiome project. These
represent problems like genome assembly and protein clustering that
until recently were done only on shared memory machines. These can now be performed
orders of magnitude faster and on data sets that were previously intractable, revealing
new species and species families.  
We see a growing number of multi-terabyte data sets but also recognize that many biologists
feel constrained in their experimental design by the daunting task of computational 
analysis. As the
demand for better performance and larger data sets continues to grow,
a distributed memory approach will be increasingly important.

Our goal in writing this paper is to summarize the work in high performance data analysis for 
genomics to help experts outside biology
understand the stress placed on parallel hardware and software systems from 
these applications.  These patterns are captured in a set of motifs, closely related to the
previous ``Seven Giants'' of data analysis, but with the critical additions of
hashing and sorting. We believe this list and the overview of application examples and 
parallelization techniques will help in designing benchmark suites,
ensuring they capture some of the most important characteristics of this
application space. The described methods can drive requirements analysis for hardware
and software, representing problems with fine-grained, asynchronous,
non-blocking, one-sided communication, irregular memory accesses, and narrow data types 
for both integers and characters.
Our experience also makes the case for reusable
software libraries that go beyond algorithms to data structures that are distributed 
across processors but can be updated by a single process with limited synchronization.